\documentclass{article}
 \font\gotb eufm10 scaled \magstep1

\newcommand{\cc}{\cite}
\newcommand{\lll}{\lambda}

\newcommand{\vp}{\varphi}
\newcommand{\R}{\hat R}
\newcommand{\I}{\hat I}
\newcommand{\Q}{\hat Q}
\newcommand{\A}{\hat A}
\newcommand{\B}{\hat B}
\newcommand {\vx}{\vp_{\xi}}
\newcommand{\PP}{\hat P}
\newcommand{\aA}{\tilde A}
\newcommand{\AAA}{\hbox{\gotb A}}

\newcommand{\BB}{\hbox{\gotb B}}
\newcommand{\sss}{\sigma}

\newcommand{\pP}{P(\aA)}
\newcommand{\bea}{\begin{eqnarray}\label}
\newcommand{\eeq}{\end{equation}}
\newcommand{\beq}{\begin{equation}\label}
\newcommand{\eea}{\end{eqnarray}}
\newcommand{\nn}{\\ \nonumber}
\newcommand{\rr}[1]{(\ref{#1})}
\newcommand {\QQQ}{\hbox{\gotb Q}_{\xi}}

\newcommand {\vpx}{\vp_{\xi}(\A)}

\begin{document}

\begin{center}
{\Large \bf Quantum mechanics as the objective local theory}  \\

\vspace{4mm}

D.A.Slavnov\\

\vspace{4mm} Department of Physics, Moscow State University,\\
GSP-2, Moscow 119992, Russia. E-mail: slavnov@theory.sinp.msu.ru\\

\end{center}

\begin{abstract}
In the work it is shown that the principles "the objective local
theory" and corollaries of the standard quantum mechanics are not
in such antagonistic inconsistency as it is usually supposed. In
the framework of algebraic approach, the postulates are formulated
which allow constructing the updated mathematical scheme of
quantum mechanics. This scheme incorporates the standard
mathematical apparatus of quantum mechanics. Simultaneously, in it
there is a mathematical object, which adequately describes
individual experiment.

\end{abstract}

In works of Bell \cc{bell1,bell2} the concept of "the objective
local theory" was introduced, in which the properties of a
physical system exist objectively, irrespective of measurement and
 are fulfilled the requirements: 1) every system is characterized by
some variables, probably, correlated for two systems;
 2) the results of measurement of one system do not depend on
 whether measurements on other system is made;
3) the characteristics of statistical ensembles depends only on
conditions in earlier times, "the retrospective causality" is
impossible.

 Actually, these requirements are the further development of
 the concept of "the physical reality", formulated in the famous work
 of Einstein-Podolsky-Rosen~\cc{epr}:  a) every element of the physical
 reality must have a counter-part in the complete physical theory;
 b) if, without in any way disturbing a system, we can predict
 with certainty (i.e. with probability equal to unity) the value
 of a physical quantity, then there is an element of physical
 reality corresponding to this physical quantity.

All these statements seem quite reasonable. However, they have one
essential drawback --- they badly agree with the basic conceptions
of the standard quantum mechanics (theory of Bohr, Heisenberg,
Dirac, von Neumann). On the other hand, the corollaries of the
standard quantum mechanics perfectly agree with enormous range of
phenomena.

In proposed work I want to show that the statements, formulated by
Bell and Einstein-Podolsky-Rosen, and corollaries of the standard
quantum mechanics are not in such antagonistic inconsistency, as
it seems at the first glance. Slightly loosening initial suppositions
of the standard quantum mechanics, not changing its corollaries,
it is possible to reconcile them with the basic principles of "the
complete objective local theory".

It is suggested to begin relaxing suppositions by  abandoning the
postulate of the standard quantum mechanics: self-adjoint linear
operators in Hilbert space correspond to observable quantities.

Instead of this supposition we shall accept much weaker postulate
of the algebraic approach to quantum theory: the elements of some
algebra corresponds to observable quantities. This phrase denotes
only that observable quantities can be summed up, multiplied
together, and multiplied by numbers. It is convenient to leave the
framework of directly observable quantities and to consider their
complex combinations, which further will be referred to as
dynamical quantities.

In this context, we adopt  {\it Postulate 1.

 Dynamic variables correspond to elements of an involutive associative
(in the general case) noncommutative algebra~\AAA, satisfying the
following conditions : \\ 1. for any element $ \R\in\AAA $  there
exists a Hermitian element  $\A$ $(\A^* = \A)$ such that $\R^*\R =
\A^2 $;\\ 2. if $\R^*\R=0 $, then $\R=0 $}.

Hereinafter, we assume that the definition of the algebra includes
the requirement that there exists a unity element~$\I$. The
observables variables correspond to the Hermitian elements of the
algebra~\AAA. We let~$\AAA_+$ denote the set of these elements.

The so-called simultaneously measurable observables, for which there
are  measuring apparatuses (system of apparatuses), permitting in
principle to measure them simultaneously to any desired degree of
precision, play a preferential role in quantum mechanics. Strictly
speaking, simultaneity is not a determinative fact. Therefore, it
is better to call these observables compatible. Peculiarity of
these observables is that for them there is a system of measuring
apparatuses which allow repeatedly to measure these observables
in an arbitrary sequence and the results for repeated measured
observables do not vary.

We adopt {\it Postulate 2} directly from the standard quantum
mechanics.

 { \it Mutually commuting elements of the set $ \AAA_+$ correspond
  to compatible observables}.

In connection with this postulate commutative subalgebras of the
algebra~\AAA{} will play essential role in the following. Let us
designate a maximal real commutative subalgebra of the
algebra~\AAA{} by~$ \QQQ $ ($ \QQQ\equiv \{\Q \}_{\xi}
\in\AAA_+$). It is algebra of the simultaneously measurable
observables.

The subscript $ \xi $ $ (\xi \in \Xi) $ distinguishes one such
subalgebra from the other. If the algebra~\AAA{} is commutative
(algebra of classical dynamical quantities), the set $ \Xi $
comprises one element. If the algebra~\AAA{} is noncommutative
(algebra of quantum dynamical quantities), the set $ \Xi $ has
potency of continuum.

The Hermitian elements of the algebra~\AAA{} are latent form of
the observable quantities. The explicit form of an observable
should be some number. This means that to determine the explicit
form of  the observables on the Hermitian elements of the
algebra~\AAA{} we must define some functional $ \vp (\A)=A $,
where $A$ is a real number.

Physically, the latent form of the observable $ \A $ becomes
explicit as a result of measurement. This means that the
functional  $\vp(\A)$ should determine the value of the observable
$\A$ that may result from {\it a concrete (individual)}
measurement. We call this functional the physical state of the
quantum object.

Only  mutually commuting elements can be measured in an individual
experiment. The sum and the product of the observables should
correspond to the sum and the product of the measurement results:
$\A_1+\A_2 \to A_1+A_2 $ and $\A_1\A_2 \to A_1A_2$.

We use the following definition. Let \BB {} be a  complex (real)
commutative algebra and $ \vp $ be a linear functional on~\BB. If
\beq{1}
 \vp (\B_1\B_2) = \vp (\B_1) \vp (\B_2)
  \eeq
 for any $\B_1, \B_2\in\BB $, then the functional $ \vp $ is called
a complex (real) homomorphism on algebra~\BB. A functional that
satisfies equality~\rr{1} also called a multiplicative functional.

 We now formulate {\it Postulate 3}.

{\it The physical state of a quantum object, appearing in the
individual measurement, is described by a (generally, multi-valued)
functional $ \vp(\A)$  $(\A\in\AAA _+)$, for which the restriction
($ \vpx $)  on any subalgebra~$\QQQ $ is single-valued and is a
real homomorphism ($ \vpx=A $  is real number)}.

 The multi-valuedness of the functional $ \vp $ is caused by the fact
 that the result of measurement can depend not only on the
 measured quantum object, but also on the nature of the measuring
 device. Let us say that the device, measuring an observable $\A$,
 is coordinated with subalgebra of observables $ \QQQ \quad (\A \in
 \QQQ) $, if for any physical state $ \vp $ the measurement
 outcome is $ \vpx $. The coordination of the device with this or
 that subalgebra $\QQQ$ is determined by its classical
 characteristics, i.e. by the construction,  position in space
 and so forth.

  Thus, the functional $ \vp(\;) $ does not describe the value of
  an observable $ \A $ in a particular physical state. It
  describes reaction of the  particular type of  measuring device to
  the observable $ \A $. Correspondingly a physical reality is not
  the value of an observable $ \A $ in the considered physical
  state, but a reaction of the measuring device to this state.

  If the functional $ \vp (\;) $ is single-valued in the point $
  \A $, we shall say that the corresponding physical state $ \vp
  $ is stable on the observable $ \A $.

It is possible to show~\cc{rud}  that the functionals, appearing
in the third postulate, have the properties:
 \bea{3}
  &/1/& \vx(0)=0; \nn {}
   &/2/& \vx (\I) =1; \nn {}
 &/3/& \vx (\A^2) \ge 0; \nn {}
  &/4/& \mbox{if } \lll =\vx (\A), \mbox{ than } \lll\in\sss (\A); \nn {}
&/5/& \mbox {if } \lll\in\sss (\A),  \mbox{ than } \lll =\vx (\A)
\mbox{ for some } \vx (\A).
 \eea
 Here $\sss(\A)$ is a spectrum of the element $\A$ in
algebra \AAA. In the standard quantum mechanics the corresponding
properties of individual measurements are postulated, here they
are consequences of the third postulate.

 Multi-valuedness of the functional $ \vp (\;) $ allows to introduce
 it in a consistent manner. It is possible to make it by direct
 construction. Due to multi-valuedness of the functional $ \vp $, the
 conditions of the Cohen-Specker no-go theorem~\cc{ksp} are not
 fulfilled for it.

It seems natural  to impose the following additional requirement
on the algebra~$\AAA$: for any two observables let there exist
experiment (i.e. a functional~$ \vp (\;) $), that separate them.
Accordingly, we formulate {\it Postulate 4:}
 $$ \vp (\A_1) = \vp (\A_2) \mbox { {\it for all} } \vp,
\mbox { {\it if and only if} } \A_1 = \A_2. $$
 In the standard quantum mechanics more strong supposition is done.

We now introduce the construction that corresponds to a pure
state in the standard quantum mechanics. The functional $ \vp $
maps the set $ \QQQ = \{\Q \}_{\xi} $ (maximal commutative
subalgebra) onto a set of real numbers:
$$\{\Q \}_{\xi}\stackrel{\vp}{\longrightarrow}\{Q=\vp (\Q)\}_{\xi}.$$

The sets $\{\vp_i (\Q) \}$, $\{\vp_j(\Q)\}$ may be different or
may coincide for different functionals $ \vp_i (\;)$, $\vp_j(\;)$.
If the relation $\vp_i(\Q)=\vp_j(\Q)=Q$ is satisfied for all
$\Q\in\{\Q\}$, then we say that the functionals $\vp_i(\;)$ and
$\vp_j(\;)$ are $\{Q\}$-equivalent. Let $ \{\vp \}_Q $ be the set
of all  $\{Q \}$-equivalent functionals, stable on the observables
of the subalgebra $\{\Q\}_{\xi}$. We call the set of the
corresponding physical states the  quantum (pure) state; it is
denoted by~$\Psi_Q $. We call quantum~$ \Psi_Q $-ensemble the set
of the physical systems that are in the quantum state~$ \Psi_Q$ .

It is not possible to obtain complete information about the
functional $\vp$ experimentally. Indeed, only mutually commuting
observables can be measured in an individual experiment. All these
observables are elements of some subalgebra~$\QQQ$. However, all
$\{Q\}$-equivalent  functionals take the same values on any
observable $\Q\in\{Q\}_{\xi}$. Therefore, the experiment cannot
help distinguish one functional~$\vp$ from the other if both
functionals belong to the set $ \{\vp \}_Q $.

Strictly speaking, the above definition of quantum state is valid
only for a physical system in which there are no identical
particles. For  description of identical particles it is necessary
to extend definition of quantum state~\cc{slav,slav1}.

The set of physical states belonging to one quantum state has
potency of continuum. Beforehand we have no special possibilities
to organize an experiment so that the preassigned physical state
appeared in it. Therefore, probability of appearance of the
investigated quantum object in a preassigned physical state is
equal to zero. It follows from this that the probability to meet
two identical physical states is equal to zero. Moreover, in
any finite or denumerable set of experiments the probability of
meeting two identical physical states is equal to zero. Certainly,
it does not signify that such event is impossible. Nevertheless,
we can consider that in different experiments we always deal with
different physical states.

Let us consider the quantum~$\Psi_Q $-ensemble as a general
population (in sense of probability theory), and any experiment
aimed to measured an observable $\A$ as a trial. Let the event
$\aA$ be experiment in which  the measured value of the observable
$\A$ is no larger then $\aA$, i.e., $\vp(\A)=A\le\aA $. This event
is not unconditional. By virtue of the second postulate one trial
cannot be event for two noncommuting observables. The probability
of the event $\aA$ is determined by structure of quantum ensemble
and this condition. Let this probability be equal to $\pP$.

We designate $\{\vp \}^{\A}_Q$ \quad
($\{\vp\}^{\A}_Q\subset\{\vp\}_Q$)  the set of the physical
states, which figure in denumerable sample of mutually independent
random trials for measurement of the observable $\A $. Let us
remark that if $\B$ is an observable, which is not commuting with
$ \A $, the probability of intersection of sets $ \{\vp \}^{\A}_Q
$ and $\{\vp\}^{\B}_Q $ is equal to zero.

Really, on the one hand, in one trial it is impossible to measure
observables $\A$ and $\B$. On the other hand, in two random
denumerable  sample a probability of reiteration of the same
physical  is equal to zero.

By definition, the probability of appearance of the event $ \aA $
in each trials is equal to $ \pP $. It determines a probability
measure $ \mu (\vp) $ $ (\vp (\A) \le \aA) $ on any such sample.
In  its turn, the measure $ \mu (\vp) $ determines distribution of
values $A_i = \vp_i (\A) $ of the observable $\A$  and expectation
$ <A> $ in this sample:
$$<A>=\int_{\{\vp\}^{\A}_Q}d\mu(\vp)\,\vp(\A).$$

 Let for $1\le i\le n$ functionals $\vp_i \in \{\vp \}^{\A}_Q $,
  then according to the theorem of Hinchin (see, for example~\cc {korn})
   the aleatory variable $ \bar A_n = (A_1 + \dots+A_n) /n $ converges on
probability to $ <A> $ as $n\to\infty $. Thus,
\beq {7}
\mbox{P-}\lim_{n\to\infty} \frac{1}{n} \Big (\vp_1 (\A) + \dots +
\vp_n (\A) \Big) = < A > \equiv \Psi_Q (\A). \eeq

Formula \rr{7} defines the functional (quantum average) on the
set $\AAA_+ $. The totality of quantum experiments leads to conclusion
that we should accept {\it Postulate 5}.

 {\it The functional $ \Psi_Q (\;) $ is linear on the set $ \AAA_+ $.}

 This means that
$$ \Psi_Q (\A + \B) = \Psi_Q (\A) + \Psi_Q (\B)$$
 even in the case where  $[\A, \B] \ne0$.

Any element $ \R $ of the algebra \AAA{} can be uniquely expressed in form $\R=\A+i\B$, where $\A,\B\in\AAA_+ $.
Therefore, the functional $\Psi_Q(\;)$ can be continued to a linear functional on the algebra \AAA: $\Psi_Q(\R)=\Psi_Q(\A)+i\Psi_Q(\B)$.

The equality $\|\R\|^2 \equiv\sup_Q \Psi_Q (\R^*\R) $ defines a
norm of an element $\R\in\AAA$. This norm has properties $\|\R^*\|
= \|\R\|$,  $\|\R^*\R\| = \|\R\|^2 $. With such norm the
algebra~\AAA{} is a $C^*$-algebra. Therefore, according to the
Gelfand-Naumark-Segal construction (see, for example,~\cc{emch}),
the functional $ \Psi_Q (\;) $ canonically generates a Hilbert
space and the representation of the algebra~\AAA{} by linear
operators in this Hilbert space. In other words, in the proposed
approach it is possible to reproduce the mathematical formalism of
the standard quantum mechanics completely.

At the same time, such only mathematical notions as vectors of
Hilbert space and operators in Hilbert space are not the primary
elements of the theory in the proposed approach. They arise only
at the second stage. The primary elements are the observables and
physical states which are related directly to the results of the
experiment. It is possible to say that they are connected
with the material structure of the investigated quantum object and
do not depend on the observer.

The physical state is "a physical reality", which Einstein did not
see in the standard quantum mechanics, and so he considered that
the quantum mechanics is the incomplete theory.

At the same time, the physical states can play a role of the
Bell's variables which characterize physical system. Here,
however, it is necessary to make an essential remark. Bell implied
that these variables are certain numerical parameters (practically
hidden parameters). On their basis  he has obtained the famous
inequality, which contradicts corollaries of the standard quantum
mechanics and experiments, that is especially essential.

In the proposed approach  physical states are nonlinear
functionals (many-valued). The Bell's inequality does not follow
from existence of such specific "hidden parameters"
(see~\cc{slav,slav2}). Besides, for such "hidden parameters" the
reasoning of von Neumann~\cc{von} on impossibility of existence of
hidden parameters is not valid (see~\cc{slav,slav1,slav2}).

Here, the basic fact is the nonlinearity of the functional
describing the physical state. Actually von Neumann has shown that
the linearity of a state is in the conflict with causality and the
hypothesis about the hidden parameters. From this he has made a
deduction that causality is absent at the microscopic level, and
the causality occurs due to averaging over large number of
noncausal events at the macroscopic level.

The approach formulated in the present work  allows to solve the
same conflict in the opposite way. It is possible to suppose that
there is causality at the level of a single microscopic
phenomenon, and the linearity is absent. The linearity of the
(quantum) state occurs due to averaging over quantum ensemble. The
transition from a single phenomenon to the quantum ensemble replaces
the initial determinism by probabilistic interpretation.

The hypothesis about the physical states allows to solve the problem of a
choice. The standard quantum mechanics considers only
probability of this or that event. At the same time, in each
concrete experiment we deal with a particular result. Who does
the corresponding choice? The God playing dice? Abstract ego?
Human consciousness? Mind of the observer?
Any of the above variants does not satisfy me.

It seems that the physical state which is determined by the
material structure of a concrete explored object allows quite
objective (and materialistic) to solve  the problem of choice.

At the same time, the proposed approach reserve to  observer
certain freedom of choice. The matter is that one physical state
can belong to different sets $ \{\vp \}_Q $ and $ \{\vp \}_P $.
Here in $ \{\vp \}_Q$ for definition of equivalence is used the set $ \{\Q
\}$ of mutually commuting observables $ \Q $, and in $
\{\vp \}_P $  are used the observables $\PP\in \{\PP \} $, where
observables $ \Q $ and $ \PP $ do not commute among themselves.
Then depending on a set ($ \{\Q \} $ or $ \{\PP \} $), i.e.
measuring apparatus, which the observer will choose (the freewill) for
classification,  the physical state $ \vp $ will be referred
either to the quantum state $ \{\vp \}_Q $ or to the quantum state
$ \{\vp \}_P $.

Thus, in contrast to the physical state, the quantum state is not
the quite objective characteristic of a quantum object. To  my
opinion,  it explains a hypertrophied role that is  ascribed
to the observer in the standard quantum mechanics.

We now discuss the place of the approach to quantum theory
proposed  in this article among different possible approaches. In
any modern approach to quantum theory, the main structure elements
are the observables and the states of the physical system. In
standard quantum mechanics, the basic structure element is the
Hilbert space. The observable variables are associated with
self-adjoint linear operators in this space, while the states are
associated with either vectors (more precisely, rays) or
statistical operators (density matrices).

The mathematical formalism constructed within the framework of
standard  quantum mechanics works very well. The question
therefore arises whether it is worthwhile to make efforts towards
constructing quantum-theoretical models beyond standard quantum
mechanics. There are good reasons for making such efforts. First
is the problem of the physical interpretation. The Hilbert space
is a rather specific mathematical concept whose physical
interpretation is not straightforward.

The consistent formulation of standard quantum mechanics relies on a special "quantum logic" that is also hard to interpret physically. All this is the subject of an everlasting debate on the consistent interpretation of quantum mechanics, a debate that intensified in recent years.

The other reason to attempt to go beyond standard quantum mechanics is the difficulties that the theory encounters in describing systems with an infinite number of degrees of freedom. As shown by von Neumann,
all representations of the canonical commutation relation for systems with a finite number of degrees of freedom are equivalent; therefore, standard quantum mechanics can provide a universal description of such
systems. However, this statement cannot be made for systems with an infinite number of degrees of freedom.

Therefore, a more general construction is needed for obtaining
their universal description. Exactly such a construction can be
realized in the algebraic approach. There are several versions of
the algebraic approach. However, the basic  concept common to a
these versions is the algebra of observables (global or local).
However, abstract this concept may seem, the algebra of
observables has a rather simple physical interpretation.

The other ingredient of the algebraic approach is the set of
linear positive functionals on the algebra of observables,
interpreted as the set of states of the physical systems. These
functionals have a rather simple physical interpretation. It is
postulated that the value of the functional on an element of the
algebra coincides with the average value of the observable for
physical systems that are in the corresponding state.

On one hand, the framework of the algebraic approach is wider than
that  of the standard quantum mechanics. Therefore, the algebraic
approach holds promise to help circumvent (at least partly) the
difficulties that standard quantum mechanics faces. On the other
hand, the Gelfand-Neimark-Segal construction allows obtaining the
principal components of standard quantum mechanics, i.e., the
Hilbert space and linear operators, within the algebraic approach.
In the framework of the algebraic approach, these concepts are not
primary but secondary. The construction used in standard quantum
mechanics is one of the possible nonequivalent representations of
the algebra of observables.

Hence, the passage from standard quantum mechanics to the
algebraic  approach is a passage to a deeper level of the quantum
theory. In this sense, the system of postulates proposed in this
article can be interpreted as a passage to a level that is yet
deeper. The first two of the formulated postulates related to the
algebra of the observables are adopted directly from the
traditional algebraic approach. Other than this, however, the
approach we propose is different.

In this approach, the principal role belongs to real commutative
subalgebras of observables. Using these subalgebras, we introduce
the essentially new element, i.e., the nonlinear functional $\vp$
whose values describe the possible result of an individual
measurement (see Postulate 3). This functional is interpreted as
the physical state. In contrast, we call the state defined in the
traditional algebraic approach (the linear functional) the quantum
state.

In the proposed approach, the primary element is the physical
state,  while the quantum state turns out to be a secondary
element, the equivalence class of the physical states. In standard
quantum mechanics and the traditional algebraic approach, the
result of an individual measurement (the physical reality)does not
have any mathematical counterpart. In the proposed approach, such
a counterpart is the nonlinear functional $\vp$.

The functional $\vp$ allows a transparent physical interpretation.
In addition, this functional allows bringing the structure of the
algebra of observables to the form of the $C^*$-algebra. This
allows using the Gelfand-Neimark-Segal construction and passing to
the mathematical scheme of standard quantum mechanics. The
traditional algebraic approach normally postulates that the
algebra of observables is the $C^*$-algebra. This is not quite
obvious physically.

Standard quantum mechanics and the traditional algebraic approach
postulate that the functional associated with the quantum state
statistically describes the average value of the corresponding
observable. In the proposed approach, this functional is
constructed through statistical averaging of the functionals
$\vp$. This construction implies that the functional describes the
average value of the observable.

\end{document}